\documentstyle[prl,aps]{revtex}

\begin{document}

\preprint{ILC/qc+dec-12dec95}

\draft

\twocolumn[

\title{Effects of Loss and Decoherence on a Simple Quantum Computer}

\author{Isaac L. Chuang$^1$, R. Laflamme$^2$, J.-P. Paz$^3$, 
		{\em and\/} Y. Yamamoto$^1$}
 
\address{\vspace*{1.2ex}
 	\hspace*{0.5ex}{$^1$ERATO Quantum Fluctuation Project} \\
 	Edward L. Ginzton Laboratory, Stanford University, 
 		Stanford, CA 94305 \\[1.2ex] 
 	$^2$Theoretical Astrophysics, T-6, MS B288	\\
 	Los Alamos National Laboratory, Los Alamos, NM 87545, USA \\[1.2ex] 
	$^3$Departamento~de~Fisica, FCEN, UBA \\
	Pabellon~1, Ciudad~Universitaria, 1428~Buenos~Aires, Argentina
}

\date{\today}
\maketitle

]


\begin{abstract}
We investigate the impact of loss (amplitude damping) and
decoherence (phase damping) on the performance of a simple quantum
computer which solves the one-bit Deutsch problem.  The components of
this machine are beamsplitters and nonlinear optical Kerr cells, but
errors primarily originate from the latter.  We develop models to
describe the effect of these errors on a quantum optical Fredkin gate.
The results are used to analyze possible error correction strategies
in a complete quantum computer.  We find that errors due to loss can
be avoided perfectly by appropriate design techniques, while
decoherence can be partially dealt with using projective error
correction.
\end{abstract}



\section{Introduction}

Quantum computers utilize the non-locality of quantum physics to allow
exponentially fast solutions to classical problems\cite{Deutsch89}.
However, there is a catch.  The fundamental element of a quantum
computer is the quantum bit (qubit), which may be in a superposition
state of zero and one.  It is a very fragile state.  Ideally, the
quantum computer is a closed system, but in reality, when information
leaks out the qubits collapse and errors are introduced into the
calculation.  Evaluation of the impact of this {\em decoherence\/}
process is a key to understanding the feasibility of quantum
computation\cite{Landauer94,Unruh95,Chuang95a,Palma95}.

In this paper we will investigate in detail the effect of decoherence
on the quantum computer of Chuang and Yamamoto\cite{Chuang94}, which
specifically implements the Deutsch-Josza solution to the one-bit
oracle problem proposed by Deutsch\cite{Deutsch92}.  The function of
their machine is essentially to use an interference experiment to
determine the class of a hidden function.  There are only two
possibilities, and in the absence of error the class is determined
with certainty.  The proposed realization uses an optical quantum
computer with beam splitters and non-linear Kerr medium
(Figure~\ref{fig:machine}).  The function of each individual component
is understood well from the study of quantum optics: the top and
bottom pairs of single mode waveguides implement two qubits, the
beamsplitter implements a $\sqrt{\mbox{\sc not}}$ logic gate, and the
Kerr medium implements a logic gate via cross-phase modulation between
single photons.

Two important imperfections which lead to errors are energy loss and
decoherence.  The former occurs due to imperfect experimental
implementation, such as facet reflections or absorption in waveguiding
media.  This loss of photons to the environment through scattering is
a distributed process which is mathematically described as {\em
amplitude damping}.

Decoherence is different; it is present even in cases in which energy
loss is negligible.  Idle photons decohere very little at room
temperature as they interact weakly with their environment.  However,
in the optical quantum computer we are analyzing the situation is
different -- the atom-photon interaction that occurs when the photons
interact with each other through a Kerr medium causes decoherence.  In
general, this interaction leaves correlations between atoms and
photons thus leading to decoherence, since partial information about
the photon's state is left behind in the atoms.  This decoherence
process is mathematically described as {\em phase damping} or phase
randomization.  The stochastic nature of decoherence likens it to a
noise process and makes it a more insidious source of errors than
loss; its time-scale may be shorter than for energy
loss\cite{Zurek91}, and as a practical matter it is harder to avoid.

We will first calculate the general effect of loss and decoherence on
an optical quantum Fredkin gate, which is a typical elementary
operation for a quantum computer.  Our models will then be used to
analyze the effects on a complete system, the Chuang-Yamamoto quantum
computer.  Finally, we shall find these results useful in
understanding how error correction can play a role in stabilizing
quantum computations.

\section{Decoherence of an optical Fredkin gate}

The apparatus used to construct an optical Fredkin
gate\cite{Yamamoto88,Milb89} is shown in Figure~\ref{fig:m-flg}.  Let
the usual creation and annihilation operators for the three modes be
$a$, $b$, and $c$ and their adjoints.  In this language, the two
beamsplitters (Figure~\ref{fig:qmbs}) are described by the operators
$B$ and ${B^\dagger}$, where
\begin{equation}
	B = \exp \left[\rule{0pt}{2.4ex}{ \frac{\pi}{4} 
		\left(\rule{0pt}{2.4ex}{ a^\dagger b - {b^\dagger} a
			}\right) }\right] 
\,.
\label{eq:bdef}
\end{equation}
Similarly, the cross-phase modulation component of the Kerr medium is
usually described by
\begin{equation}
	K = \exp \left[\rule{0pt}{2.4ex}{ i\chi \, {b^\dagger b}
			\,{c^\dagger c} }\right] 
\,,
\end{equation}
with $\chi=\pi$ ideally.  These definitions immediately give us the
quantum Fredkin gate operator (with no damping and no decoherence), 
\begin{equation}
	F = {B^\dagger}\, K \, B
\,.
\end{equation}
The matrix elements of $F$ relevant for the Chuang-Yamamoto quantum
computer, to be discussed later, are
\begin{eqnarray}
	F \, |{000}\rangle  &=&   |{000}\rangle 
\\
	F \, |{100}\rangle  &=&   |{100}\rangle 
\\
	F \, |{010}\rangle  &=&   |{010}\rangle 
\\
	F \, |{101}\rangle  &=&   |{011}\rangle 
\\
	F \, |{011}\rangle  &=&   |{101}\rangle 
\,,
\end{eqnarray}
using the labeling $|{\sf abc}\rangle $.  Note that ${F^\dagger} = F$.  

Energy loss, i.e. the loss of a photon to the environment, can be
described for a single qubit (in mode {\sf a}) by the superscattering
operator $\$_{\gamma}^a$
\begin{equation}
	\$_\gamma^a 
	   \left[
	    \begin{array}{cc} 
		\rho_{00}  & \rho_{01} \\
		\rho_{10}  & \rho_{11} \\
	    \end{array}
	    \right]
	  = 
	   \left[
	    \begin{array}{cc} 
		  \rho_{00}  + (1-e^{-\gamma}) \rho_{11} 
		& e^{-\gamma/2} \rho_{01} \\
		  e^{-\gamma/2} \rho_{10}  
		& e^{-\gamma}   \rho_{11} \\
	    \end{array}
	    \right]
\,.
\label{eq:damping}
\end{equation}
A simple way of deriving this expression is by taking $1-e^{-\gamma}$
as the probability for absorbing the qubit photon and creating an
excitation in the unobserved environment.  Mathematically, we may
write the wavefuntion for the qubit + environment as
\begin{equation}
	|{10}\rangle \rightarrow 
		e^{-\gamma/2}|{10}\rangle  + \sqrt{1-e^{-\gamma}} |{01}\rangle
\,.
\end{equation}
and arrive at the superscattering operator by summing over the
environment, represented by the second label.  Thus, $\$_{\gamma}^a$
describes amplitude damping due to coupling of a qubit to its
environment.

The Kerr medium used in the quantum Fredkin gate is experimentally
known to be lossy\cite{Watanabe90}, and we may model this by inserting
a loss mechanism in its arguments.  The resulting Fredkin gate is
described by the superscattering operator $\$_{F_\gamma} = {B^\dagger}
\$_{\gamma}^b \$_{\gamma}^c K B$.  (It can be shown that the physics
does not change if the damping is distributed or placed before or
after the Kerr medium.)  It is clear that the worst affected states
for the simple quantum computer are $|{101}\rangle $ and
$|{011}\rangle $ for which
\begin{eqnarray}
	\$_{F_\gamma} \left[\rule{0pt}{2.4ex}{ |{101}\rangle
			\langle{101}| }\right] 
	&=& 
         \frac{(1-e^{-\gamma})^2}{2} |{000}\rangle \langle{000}| 
\\
	&& \hspace*{-11ex} +\ 
         	\frac{e^{-\gamma}(1-e^{-\gamma})}{2} |{001}\rangle
			\langle{001}| 
\nonumber
\\
	&& \hspace*{-11ex} +\ 
		\frac{(1-e^{-\gamma})}{4} |{\phi_{01}\rangle
				}\langle{\phi_{01}|} 
		+\, \frac{e^{-\gamma}}{4} |{\phi_{10}\rangle
				}\langle{\phi_{10}|} 
\,,
\label{eq:adampfg}
\end{eqnarray}
where
\begin{eqnarray}
	\phi_{01} &=& (1+e^{-\gamma/2})\, |{010}\rangle 
				+ (1-e^{-\gamma/2}) \, |{100}\rangle 
\\
	\phi_{10} &=& (1+e^{-\gamma/2})\, |{011}\rangle  
				+ (1-e^{-\gamma/2}) \, |{101}\rangle 
\,.
\end{eqnarray}
The first term in $\$_{F_\gamma}$ corresponds to the absorption of two
photons in the Kerr medium. The second and third terms result from the
absorption of one photon (from either mode {\sf b} or {\sf c}) and
finally the last term correspond to no absorption at all in the Kerr
medium.  A similar result is obtained for $\$_{F_\gamma}(|{011}\rangle
\langle{011}|)$ by interchanging the first two qubits.  Obviously
energy loss induces errors by degrading the computer's state and
transforming it into one with lower energy.  However, we will see
later how errors coming from energy loss can easily be detected and
corrected.

Even if one is able to minimize energy losses, there will be another
source of problems: damping of phase coherence which occurs as the
photons go through the Kerr medium.  In fact, the effective operator
for the Kerr cell may be written as
\begin{equation}
	K(\epsilon) = \exp \left[\rule{0pt}{2.4ex}{i\pi \, {b^\dagger
			b} \,{c^\dagger c} + i\eta}\right] 
\,,
\label{eq:kdec}
\end{equation}
where $\eta$ is an operator acting both on the photons and the atoms.
This interaction produces correlations which, as the atoms are
unobserved, generate decoherence.  For the purpose of computing the
trace over the environment we can treat $\eta$ as if it were a simple
function of a random variable $\epsilon$.  The source for randomness
is, as described above, the atom--photon interaction taking place in
the Kerr cell. We let
\begin{equation}
	\eta = \epsilon\, ({b^\dagger b} + {c^\dagger c})
\,,
\label{eq:noiseone}
\end{equation}
where $\epsilon$ is a random variable which zero mean (this will be
justified in an explicit model later).  Taking this into account, the
relevant matrix elements of the quantum Fredkin gate with phase
decoherence, $F_\lambda$, are
\begin{eqnarray}
	F_\lambda |{000}\rangle  &=&   |{000}\rangle 
\label{eq:flambdaone}
\\
	F_\lambda |{100}\rangle  
		&=&   \left[\rule{0pt}{2.4ex}{
			\frac{1+e^{i\epsilon}}{2} }\right] 
				|{100}\rangle 
			+ \left[\rule{0pt}{2.4ex}{
				\frac{1-e^{i\epsilon}}{2} }\right] 
				|{010}\rangle 
\\
	F_\lambda |{010}\rangle  
		&=&   \left[\rule{0pt}{2.4ex}{
				\frac{1-e^{i\epsilon}}{2} }\right] 
				|{100}\rangle 
				+ \left[\rule{0pt}{2.4ex}{
				\frac{1+e^{i\epsilon}}{2} }\right]	 
				|{010}\rangle 
\\
	F_\lambda |{101}\rangle  
		&=&   \frac{e^{i\epsilon}}{2} 
				\left[\rule{0pt}{2.4ex}{
				1-e^{i\epsilon} }\right] 
				|{101}\rangle 
			+ \frac{e^{i\epsilon}}{2} \left[\rule{0pt}{2.4ex}{
				1+e^{i\epsilon} }\right] |{011}\rangle 
\\
	F_\lambda |{011}\rangle  
		&=&   \frac{e^{i\epsilon}}{2} 
				\left[\rule{0pt}{2.4ex}{
				1+e^{i\epsilon} }\right] 
				|{101}\rangle 
			+ \frac{e^{i\epsilon}}{2} \left[\rule{0pt}{2.4ex}{
				1-e^{i\epsilon} }\right] |{011}\rangle  
\,,
\label{eq:flambdaN}
\end{eqnarray}
Tracing over the environment formed by the atoms corresponds to
averaging over the random variable $\epsilon$. Assuming a Gaussian
distribution, i.e. $\langle e^{i\epsilon} \rangle = e^{-\lambda}$, with
$\lambda=\langle \epsilon^2 \rangle$, one gets a superscattering operator,
$\$_{F_\lambda}$, for the Fredkin gate.  For example, its action upon
the state $|{011}\rangle $ is
\begin{equation}
	\$_{F_\lambda}\left[\rule{0pt}{2.4ex}{ |{101}\rangle
			\langle{101}| }\right] 
	= \frac{1+e^{-\lambda}}{2}|{011}\rangle \langle{011}|+
		\frac{1-e^{-\lambda}}{2} |{101}\rangle \langle{101}|
\label{eq:kerrdec}
\,,
\end{equation}
Similarly, $\$_{F_\lambda}(|{011}\rangle \langle{011}|)$ is obtained
by switching the first two bits in the previous expression.

In a recent paper Boivin et al.\cite{Boivin94} presented a model for a
1+1 dimensional Kerr cell with weak nonlinearity. The main conclusion
of their analysis, based on the fact that the variable $\eta$ in
(\ref{eq:kdec}) can indeed be described by equation
(\ref{eq:noiseone}), is that non-linearity of the medium is
unavoidably accompanied by phase noise of the field.  Using their
conclusions here implies the existence of a relationship between the
amount of phase shift $\theta$ and the value of $\lambda$.  In their
example, for a coherent input state corresponding to a monochromatic
pump at the carrier frequency, one gets that for a phase shift of
$\theta=\pi$, the decoherence parameter $\lambda = \pi\Omega/I$, where
$\Omega$ is the resonant frequency of the medium and $I$ is the
intensity of the pulse, in units of photon number per second.  This
would imply a rather large amount of decoherence per step of the
quantum computer.  It remains to be seen if their 1+1 dimensional
model is reasonable.

\section{Energy Loss in the simple quantum computer}

Let us now apply our results to investigate the effect of energy loss
and decoherence on the Chuang-Yamamoto computer.  We begin by
considering the effect of energy loss.  The computer has two
nontrivial settings, $k_1=0$ and $k_1=1$; when $k_1=0$, the unitary
transform performed by the computer is
\begin{equation}
	U_0 = {B^\dagger}_{cd} \$_{F_\gamma}^{abe} S_a F_{abe} B_{cd}
\,,
\label{eq:uzero}
\end{equation} 
where {\sf a,b,c,d} denote the four optical modes used in the machine
(mode {\sf e} is always zero).  We consider only the effect of loss in
the second Fredkin gate by replacing the operator $F_{abc}$ with the
non-unitary superscattering operator $\$_{F_\gamma}^{abc}$ which
describes a Fredkin gate with loss. The normal input to the machine is
$|{\sf abcd}\rangle = |{0101}\rangle $, and {\sf e} is the vacuum.
Since the signal in modes {\sf c} and {\sf d} do not enter the Kerr
medium when $k_1=0$, loss in the Kerr medium is irrelevant, and the
answer either $|0101\rangle$ or $|{0001}\rangle $.  The latter case is
an error but it is easily detectable as only one photon is observed
and thus must be incorrect (under proper operation, no photon is ever
lost from the system).

On the other hand, when $k_1k_0=10$, then the transform performed by
the computer is
\begin{equation}
	U_1 = {B^\dagger}_{cd} \$_{F_\gamma}^{abc} S_a F_{abc} B_{cd}
\,,
\label{eq:uone}
\end{equation}
such that for the first half of the apparatus, we have
\begin{eqnarray}
	\mbox{$|\psi_0\rangle$} &=& |0101\rangle
\nonumber\\
	\mbox{$|\psi_1\rangle$} &=& B_{cd}\mbox{$|\psi_0\rangle$} 
	= \frac{1}{\sqrt{2}} \left[\rule{0pt}{2.4ex}{ |{0101}\rangle
			+ |{0110}\rangle  }\right] 
\nonumber\\
	\mbox{$|\psi_2\rangle$} &=& F_{abc} \mbox{$|\psi_1\rangle$} 
	= \frac{1}{\sqrt{2}} \left[\rule{0pt}{2.4ex}{ |{0101}\rangle
		+ |{1010}\rangle  }\right] 
\nonumber\\
	\mbox{$|\psi_3\rangle$} &=& S\mbox{$|\psi_2\rangle$} 
	= \frac{1}{\sqrt{2}} \left[\rule{0pt}{2.4ex}{ |{0101}\rangle
		- |{1010}\rangle  }\right] 
\end{eqnarray}
as the state before the second Fredkin gate.  Using
a density matrix description, we calculate
\begin{eqnarray}
	\rho_{4} &=& \$_{F_\gamma}^{abc} \left[\rule{0pt}{2.4ex}{
		|{\psi_3}\rangle \langle{\psi_3}| }\right] 
\\
	\rho_5 &=& B_{cd}^\dagger \rho_{4} B_{cd}
\,,
\end{eqnarray}
where $\rho_{4}$ is the output of the second, lossy Fredkin gate, and
$\rho_5$ is the final output.  The diagonal elements of $\rho_5$ give
us the final measurement result probabilities.  Errors occur because
of imperfect switching, as described by Eq.(\ref{eq:adampfg}).
Following \cite{Chuang94}, if the measurement of mode {\sf d} is taken
as the computation result, we find that the error probability is
\begin{equation}
	P_{\mbox{\sc noec}} =
	\frac{1}{4} 
	\left[\rule{0pt}{2.4ex}{
					1 + e^{-\gamma} - 2e^{-3\gamma/2} 
	}\right]
\,.
\end{equation}
However, the technique of using two modes to represent a single qubit
(what we have termed the {\em dual-rail quantum bit}) allows for a
simple error correction scheme; that is, for each of the pairs {\{\sf
a,b\}} and {\{\sf c,d\}}, the only permissible states are
$|{01}\rangle $ and $|{10}\rangle $.  The states $|{00}\rangle $ and
$|{11}\rangle $ correspond to loosing or gaining a photon, which will
only happen when an error occurs.  Thus, if we reject all such illegal
results, we find that the error probability is
\begin{equation}
	P_{\mbox{\sc ec}} =
	\frac{1}{2} 
	\left[\rule{0pt}{2.4ex}{
					1 - {\rm sech}\, \frac{\gamma}{2}
	}\right]
\,,
\end{equation}
that is, just the relative probability of finding $|0101\rangle$ (the
wrong answer) to $|0110\rangle$ (the right answer).  The improvement
in error probability given by use of this simple-minded qubit error
correction scheme is shown in Figure~\ref{fig:biterror}.  

Even more interesting is what happens when the loss is {\em balanced}
such that all four modes suffer identically.  That is, we let
$\$'_{F_\gamma} = {B^\dagger} \$_{\gamma}^a \$_{\gamma}^b \$_{\gamma}^c
\$_{\gamma}^d K B$, as shown in Figure~\ref{fig:equalloss}.  Using
this in $U_1$, we find that the diagonal elements of the final density
matrix are
\begin{eqnarray}
	{\rho'}_5^{\rm diag} 
	&=&   e^{-4\gamma} |{0110}\rangle \langle{0110}|
		+ \left( 1 + {e^{-4\gamma}} - 2 e^{-2\gamma} \right) \, 
					    	|{0000}\rangle \langle{0000}| 
\nonumber
\\
	&& \hspace*{3ex} + \ 
		\frac{e^{-2\gamma}-e^{-4\gamma}}{2} 
			\left[\rule{0pt}{2.4ex}{ |{0001}\rangle \langle{0001}|
			      + |{0010}\rangle \langle{0010}|
}\right.
\nonumber \\
&&\hspace*{13ex}
\left.{
			      + |{0100}\rangle \langle{0100}|
			      + |{1000}\rangle \langle{1000}|
					}\rule{0pt}{2.4ex}\right]    
\,.
\end{eqnarray}
%
%
Furthermore, since the only legal state which can be obtained from the
above is $|{0110}\rangle $ (there must be two photons in the output), we
find that the after error correction, the error probability is {\em
zero}.  Physically, this occurs because of the symmetry of the
damping.  In classical optics, it is well-known that by balancing the
loss in an interferometer, unit visibility can be obtained.
Analogously, for a single-photon interferometer (when only one photon
is present in both arms), either the photon is lost (in which case the
output is $|{00}\rangle $), or coherence is preserved perfectly.  This
behavior is the basis for the regenerative properties of the dual-rail
quantum bit\cite{Chuang95b}.

\section{Decoherence in the simple quantum computer}

The above calculation indicates that errors due to loss can be
prevented by using the appropriate design.  However, the effect of
phase damping is more insidious.  To see this, let us substitute our
results for the noisy Fredkin gate $\$_{F_{\lambda}}$ into $U_0$ and
$U_1$ and calculate the output state $\rho_5$, just as before.  Using
Eqs.(\ref{eq:flambdaone}-\ref{eq:flambdaN}) and averaging over
$\epsilon$, we find that for the $k_1=0$ configuration the diagonal
elements of the output density matrix are
\begin{eqnarray}
	\rho_5^{\rm diag}(k_1=0) 
	&=& {\rm diag}\left[{ U'_0 \, |{0101}\rangle \langle{0101}| \,
		{U'}^\dagger_0 }\right] 
\\
	&=& \frac{1+e^{-2\lambda}}{2}\,|{0101}\rangle\langle{0101}| 
\nonumber
\\ &&\hspace*{3ex}
	       +\ \frac{1-e^{-2\lambda}}{2}\,|{1001}\rangle\langle{1001}|
\,,
\end{eqnarray}
Note that $\lambda$ parameterizes the amount of decoherence, and for
large $\lambda$, the two states $|{0101}\rangle\langle{0101}|$ and
$|{1001}\rangle\langle{1001}|$ are equally probable.  This mixed state
results because decoherence in the Kerr media performs a partial
``which path'' measurement on the interferometer formed by modes {\sf
a} and {\sf b} in the Fredkin gate.

On the other hand, when $k_1=1$, we use $U_1'$ to find the 
final result,
\begin{eqnarray}
	\rho_5^{\rm diag} (k_1=1)
	&=& {\rm diag}\left[{ U'_1 \, |{0101}\rangle \langle{0101}| \,
			{U'}^\dagger_1 }\right]
\\
		 &=&  \frac{(1-e^{-2\lambda})}{4}\,
			|{0101}\rangle\langle{0101}| 
\nonumber \\ 
&&\hspace*{3ex}
			+\ \frac{(1-e^{-2\lambda})}{4}\,
			|{1010}\rangle\langle{1010}| 
\nonumber \\
&&\hspace*{3ex}
			+\ \frac{(1+3e^{-2\lambda})}{4}\,
			|{0110}\rangle\langle{0110}| 
\nonumber \\
&&\hspace*{3ex}
			+\ \frac{(1-e^{-2\lambda})^2}{4}\,
			|{1001}\rangle\langle{1001}| 
\,.
\end{eqnarray}
In the limit of large $\lambda$, the four states
$|{0101}\rangle\langle{0101}|$, $|{1010}\rangle\langle{1010}|$,
$|{0110}\rangle\langle{0110}|$ and $|{1001}\rangle\langle{1001}|$ are
equally probable. This means that our simple-minded error correction
scheme (i.e. simply rejecting illegal states) fails!  However, we have
a more sophisticated method at our disposal.

If we have {\em a priori} knowledge that under perfect operation, the
state $|{\phi}\rangle $ after the first Fredkin gate will be either
$[|{0101}\rangle +|{0110}\rangle ]/\sqrt{2}$ or $[|{0101}\rangle
+|{1010}\rangle ]/\sqrt{2}$, then we know that the space of legal
results is spanned by $|{\psi_0}\rangle = [|{0101}\rangle
+|{1010}\rangle ]/\sqrt{2}$ and $|{\psi_1}\rangle = [|{0101}\rangle
+2|{0110}\rangle -|{1010}\rangle ]/\sqrt{6}$.  We may thus detect
errors by measuring the component of $|{\phi}\rangle $ perpendicular
to the $\{|{\psi_0}\rangle ,|{\psi_1}\rangle \}$ space.  The quantum
circuit to do this is straightforward to design; basically, we perform
a unitary transform to get $|{\phi'}\rangle = U\,|{\phi}\rangle $
which is either $|{0101}\rangle $ or $|{1001}\rangle $.  When the last
two labels are measured to be other than $|{01}\rangle $ we know an
error has occurred, and the trial is rejected.  Otherwise, we perform
the inverse transform to restore the state, and continue as before.
Using this scheme, we find that the probability of error in the final
result decreases from $\lambda-\lambda^2$ to $11\lambda/18 -
47\lambda^2/162$ for small $\lambda$.  The results are plotted in
Figure~\ref{fig:laferr}.

The improvement achieved by the above indicates the possibility of
using projective techniques to correct for phase randomization.
Ideally, it would be nice to be able to detect and correct for errors
due to decoherence, just as is possible for amplitude damping using
dual-rail qubits.  Along these lines, we have recently discovered a
qubit representation which shows significant resistance to phase
randomization\cite{Chuang95c}; instead of decohering at rate
$\lambda$, we can achieve $4\lambda/N$ for arbitrary $N$ by
introducing $N-1$ ancilla qubits (which also decohere!) appropriately
entangled.

\section{Conclusion}

Our analysis of the effects of loss and decoherence on a simple
quantum computer indicates that although decoherence is a significant
impediment to the realization of quantum computers, techniques exist
which may be utilized to mitigate errors.  In particular, the
dual-rail quantum bit representation may be used to perfectly detect
and correct errors due to amplitude damping.  Other representations
also exist which may be used against phase
randomization\cite{Chuang95c,Chuang95isqm}.

Alternatively, new technologies may be developed which allow single
photon qubits to interact without decoherence; the model used here is
based on our understanding of bulk nonlinear optical materials, but
other possibilities exist for resonant interactions which should have
high $\chi^{(3)}$, negligible loss, and short interaction lengths.
These are based on single-photonics technologies\cite{Imamoglu94} that
take advantage of our ability to engineer semiconductor devices.  For
example, we may envision a system consisting of the transmission of
single photon dual-rail qubits over a fiber-optic link, with coding,
decoding, and regeneration using exciton-polariton quantum logic gates
and high-efficiency single-photon detectors.

 


\onecolumn

\begin{figure}[p]
\caption{The Chuang-Yamamoto quantum computer used to solve the one-bit
	 Deutsch problem\protect\cite{Deutsch92}.  The apparatus in
	the dashed box is used by 
	Bob to calculate $f_k(x)$, and everything else belongs to
	Alice.  $k_0$ and $k_1$ control classical switches.
	Computation flows from left to right.}
\label{fig:machine}
\end{figure}

\begin{figure}[p]
\caption{A quantum-optical Fredkin gate constructed using a nonlinear
	Mach-Zehnder interferometer and cross-phase modulation via the Kerr
	interaction.  The beamsplitter on the left (right) is described by $B$
	(${B^\dagger})$.}
\label{fig:m-flg}
\end{figure}

\begin{figure}[p]
\caption{Classical transform functions for the 50/50 beamsplitter
	which are consistent Eq.(\protect\ref{eq:bdef}).  Note that
	$Ba{B^\dagger}=(a-b)/\protect\sqrt{2}$, and
	$Bb{B^\dagger}=(a+b)/\protect\sqrt{2}$.}
\label{fig:qmbs}
\end{figure}

\begin{figure}[p]
\caption{Model of quantum Fredkin gate with equal loss in all four modes.}
\label{fig:equalloss}
\end{figure}

\begin{figure}[p]
\caption{Error probability for the final measurement result in the
	$k_1k_0=10$ case, with and without error correction (lower and upper
	curves).  For small $\gamma$, the improvement is
	substantial; $P_{\mbox{\sc noec}}\sim
	\gamma/2$ and $P_{\mbox{\sc ec}}\sim \gamma^2/16$, where loss
	is $10 \gamma \log_{10} e$ [dB].}
\label{fig:biterror}
\end{figure}

\begin{figure}[p]
\caption{Error probability for the final measurement result using a
	projective phase decoherence error correction scheme.  The
	amount of damping is $10 \lambda \log_{10} e $ [dB].}
\label{fig:laferr}
\end{figure}

\end{document}